\documentstyle[aps,amsfonts,prl,twocolumn,floats]{revtex}

\begin{document}
\unitlength 1mm
\title{Comment on the paper `Inelastic granular gas: Computer simulations and kinetic theory of the cooling state' by J.M. Salazar and L. Brenig}
\author{Yves Elskens\cite{bylineYE}}
\address{Equipe turbulence plasma, case 321, \\
         Laboratoire de physique des interactions ioniques et mol\'eculaires,
         UMR 6633 CNRS--Universit{\'e} de Provence,
              \\
              Centre de Saint-J{\'e}r{\^o}me, av. escadrille Normandie-Niemen,
              F-13397 Marseille Cedex 20}
\date{preprint TP99.14}
\maketitle
\begin{abstract}
A hierarchy of equations was introduced recently to describe the
kinetics of homogeneous cooling of the gas of inelastic hard
spheres. It is argued that this hierarchy does not describe this
system accurately, and a simple test for this is proposed.
A stochastic interpretation of the hierarchy is presented.\\%
\vskip0.5cm
PACS numbers: \\%
83.70.Fn Granular solids \\%
05.20.Dd Kinetic theory
\end{abstract}
\pacs{PACS numbers: 83.70.Fn, 05.20.Dd}

\narrowtext

In their recent papers \cite{BS1,BS2}, Brenig and Salazar
investigate the kinetic theory of inelastic hard spheres. They
obtain a hierarchy of ordinary differential equations for the
moments $x_k = \langle W^k \rangle = \langle |{\hat {\bf r}}_{12}
\cdot {\bf v}_{12}|^k \rangle$ of the modulus $W$ of the radial
component of the relative velocity of two particle in the
homogeneous cooling regime. Their equations are remarkably simple
and may shed new light on this fundamental process of granular
media.

It seems however that their derivation is incorrect.
This Comment aims at (1) introducing a probabilistic interpretation of
their equations, (2) arguing on the core of the illegitimate
derivation and (3) on a first-principle problem raised by it, (4)
commenting on processes for which their equations may hold and (5)
proposing a simple test for these equations.

{\bf 1.} The Brenig-Salazar hierarchy reads for all $k \geq 1$
\begin{equation}
  \dot x_k = - (1 - \alpha^k) x_{k+1}
\label{xk}
\end{equation}
with the dot denoting derivation with respect to rescaled
\cite{dim} time $t = t'/\tau$, where $t'$ is the physical time,
and $\tau$ is the characteristic intercollision distance. Given
the spheres diameter $a$ and number density $n$, $\tau = (2 \pi
a^2 n)^{-1}$ in three space dimensions and $\tau = (\pi a n)^{-1}$
in two space dimensions. The dimensionless parameter $0 \leq
\alpha \leq 1$ is the (normal velocity) restitution factor at the
collision.

Denoting by $f(w,t)$ the probability distribution function of $W$,
hierarchy (\ref{xk}) is meaningful provided all moments $x_k (t) =
\langle W^k \rangle = \int_0^\infty w^k f(w,t) dw$ are finite.
Then $f(w,t)$ itself is uniquely determined \cite{moments} by the
moments $x_k(t)$. Elementary algebra rewrites (\ref{xk}) as
\begin{equation}
  \partial_t f(w,t) = - w f(w,t) + \alpha^{-2}w f(\alpha^{-1}w,t)
\label{dft}
\end{equation}
The latter equation admits a natural interpretation as a random
walk over $[0,+\infty[$ for the normal relative velocity $W$ of a
pair of particles : for $\Delta t \to 0$, if $W(t) = w$, it may
jump to $W(t + \Delta t) = \alpha w$ with probability $w \Delta t$
during the time interval $\Delta t$, and it may remain $W(t +
\Delta t) = w$ with probability $1 - w \Delta t$. Thus successive
velocity jumps (collisions) occur independently, with a waiting
time between them distributed according to an exponential law
(with parameter $W(t)$).

{\bf 2.} Following notations of the appendix of Ref. \cite{BS2},
this stochastic process describes indeed the result of inelastic
collisions between particles 1 and 2.

However both particles are surrounded by other particles, and a
collision of (say) particle 1 with a third particle 3 will change
${\bf v}_{12}$ to some ${\bf v}'_{12}$ : in general $|{\hat {\bf
r}}_{12} \cdot {\bf v}_{12}| \neq |{\hat {\bf r}}_{12} \cdot {\bf
v}'_{12}|$. Salazar and Brenig claim that the balance of these
processes vanishes. This is doubful mathematically and physically.

The mathematical difficulty originates from the Heaviside
functions $\Theta(\pm g_{13})$ in (A7) and from the fact that
$|g'_{12}| \neq |g_{12}|$, $|g'_{13}| \neq |g_{13}|$. The two
terms in the integrand cannot be reduced to each other, thus only
a deeper symmetry can yield $K_{123}=0$. The alleged symmetry,
that expression (A10) be odd in ${\hat {\bf r}}_{13}$, overlooks
the collisional requirement that $g'_{13}<0$ \cite{signs} : (A10)
contributes to (A7) only for a half-space integration, with
constant sign. The other half-space integration also contributes
with a constant sign, with a different integrand. Nothing warrants
a cancellation.

Physically, for $\alpha = 1$, in a homogeneous stationary state,
detailed collisional balance ensures that processes $13 \to 1'3'$
have rates equal to processes $1'3' \to 13$. But such a detailed
balance does not apply to inelastic collisions -- even in a
homogeneous cooling regime. \cite{distfn}

{\bf 3.} From a more general viewpoint, the probability
distribution function of ${\bf v}'_{12}$ depends not only on
velocities ${\bf v}_1$, ${\bf v}_2$ and ${\bf v}_3$ but also on
the distribution of the radial vector at impact ${\hat {\bf
r}}_{13}$. Assuming that the distribution of the latter is
continuous, the projection $W' = |{\hat {\bf r}}_{12} \cdot {\bf
v}'_{12}|$ after collision 13 has a continuous probability
distribution~; one must even allow for $W' > W$ since particle 3
may hit particle 1 on any side. Third-party collisions thus
contribute to $\partial_t f$ an integral kernel which is
continuous (rather than the Dirac distribution $\delta(w_{\rm in}
- \alpha^{-1} w_{\rm out})$ of (\ref{dft})). Now, the right hand
side of (\ref{dft}) is a sum of terms describing the various
collisions, each with {\em positive} kernel. Once a process has
been allowed, no competing process can compensate it
statistically~: averaging with other processes acts in a convex
manner on the infinitesimal generators of the random process.

Equation (\ref{dft}) accounts neither for processes increasing $W$
nor for processes smoothing a discrete initial distribution of
$W$. Therefore is appears unable to describe accurately the kinetics of
the gas of inelastic hard spheres.

Adding to the right hand side
of (\ref{dft}) a diffusion term \cite{RL} seems a promising way to
include the effect of third party collisions, though an integral
kernel is more general. Note that, for $N \gg 1$, collisions 13
and 23 are ${\cal O}(N)$ more frequent than direct collisions 12,
so that their contribution would be diffusion-like by a central
limit effect and could overwhelm the 12 term taken into account in
(\ref{dft}).

{\bf 4.} Does hierarchy (\ref{xk}) model the inelastic Lorentz
gas~? In this case, an ensemble of independent light particles
move ballistically between collisions with fixed, infinitely heavy
scattering obstacles~; the normal component of the light
particle's momentum is multiplied by $\alpha$ at each collision.
Again, the outgoing normal velocity of the light particle after
one collision is generally not its incoming normal velocity for
the next collision, as successive normal vectors are not parallel,
hence equations (\ref{xk}) and (\ref{dft}) do not apply. Moreover,
with fixed scatterers of given size, the light particle may
undergo a sequence of correlated collisions, while equations
(\ref{xk}) and (\ref{dft}) describe independent collisions.

There may however be a limit in which the differential equation
(\ref{dft}) and hierarchy (\ref{xk}) could describe the evolution
of a light particle velocity. This limit is a memoryless
stochastic model, in which scatterers are Poisson distributed in
space, the modulus (instead of normal component) of the light
particle outgoing velocity is $\alpha$ times its incoming velocity
modulus, and the outgoing direction is chosen so that recollisions
would have vanishing probability in a Grad limit (scatterer
density $n \to \infty$, scatterer size $a \to 0$, and $\tau$
constant). This model is not a limit of the previous Lorentz gas.

{\bf 5.} Finally, independently from the derivation of the
hierarchy from the $N$-body Liouville equation \cite{BS2}, one may
also test the relevance of hierarchy (\ref{xk}) and equation
(\ref{dft}) to homogeneous cooling (or to other physical
processes) directly through an explicit prediction. For
kinetics described by (\ref{dft}), all expectations
\begin{eqnarray}
  c_n (t) &=& \int_0^\infty \cos (2 \pi n \ln(\kappa w) / \ln \alpha) f(w,t) dw
  \\
  s_n (t) &=& \int_0^\infty \sin (2 \pi n \ln(\kappa w) / \ln \alpha) f(w,t) dw
\end{eqnarray}
are constant in time, where an arbitrary constant $\kappa$ is
introduced for dimensional reasons. Indeed, the random walk
associated with (\ref{dft}) changes $\ln (\kappa W(t)) / \ln
\alpha$ by unit steps $+1$, leaving the trigonometric
functions invariant.

It is a pleasure to thank L. Brenig, R. Lambiotte and J.M. Salazar
Cruz for discussions, and G. Nicolis for his invitation to
universit{\'e} libre de Bruxelles which initiated these
discussions.


\clearpage


\begin{references}

\bibitem[*]{bylineYE} {Email : elskens@newsup.univ-mrs.fr}

\bibitem{BS1} {L. Brenig and J.M. Salazar, J. Plasma Phys. {\bf 59}, 639 (1998).}

\bibitem{BS2} {J.M. Salazar and L. Brenig, Phys. Rev. E {\bf 59}, 2093 (1999).}

\bibitem{dim} {The reduced time is dimensionally equivalent to the reciprocal
of a velocity, as rescaling all initial velocities leaves the
dynamics invariant, up to the time scale. This and further
dimensional analysis hints that, whatever the exact equations
obeyed by the moments, algebraic scalings $x_k(t) \sim t^{-k}$
should hold for $t \to \infty$. }

\bibitem{moments} {Use the Fourier transform representation
$\tilde f(z,t) = \langle e^{i z W} \rangle = \sum_{k=0}^\infty
x_k(t) (i z)^k / k!$.}

\bibitem{signs} {The Heaviside functions in collision kernels in the appendix
seem inconsistent with the definition $g_{12} = {\hat {\bf
r}}_{12} \cdot {\bf v}_{12}$, which is positive for particles
moving away from each other. This sign reversal does not
affect our discussion.}

\bibitem{distfn} {An interesting motivation for establishing
equations (\ref{xk}) and (\ref{dft}) is to deduce from them the
moments $x_k$ in the homogeneous cooling regime.
Indeed, a scaling form of $f$ and corresponding
time-dependent moments can be constructed analytically
\cite{Elskens}. For comparison, in the elastic case $\alpha=1$,
all moments and $f$ are constant in time according to (\ref{xk})
and (\ref{dft}), and apparently arbitrary~;
it is Boltzmann's analysis of the collisions which enables one
to identify the Maxwell-Boltzmann distribution for ${\bf v}_{12}$
in the homogeneous stationary state (from which relations
between the moments can be deduced).}

\bibitem{Elskens} {Y. Elskens, in {\it Nonlinear science :
dynamics and stochasticity}, to be published.}

\bibitem{RL} {R. Lambiotte, private communication.}

\end{references}
\end{document}